\documentclass[preprint]{aastex}

\slugcomment{{\bf AJ, accepted 25 Jan 2008 }}

\shorttitle{NGC 6397}
\shortauthors{Anderson}

\def\minspt{$\buildrel{\prime}\over .$}
\def\secspt{$\buildrel{\prime\prime}\over .$}
\newcommand{\hst}{{\it HST}}

\renewcommand{\footnoterule}{\null}

\begin{document}
\vskip-0.7truecm

\title{A Deep \hst\ Study of the Globular Cluster NGC 6397:  Reduction
Methods\footnote{Based  on  observations with  the  NASA/ESA {\it Hubble
Space Telescope},  obtained at the Space  Telescope Science Institute,
which is operated by AURA, Inc., under NASA contract NAS 5-26555.}}

\author{Jay Anderson\altaffilmark{2}, 
        Ivan R.\ King\altaffilmark{3},
        Harvey B.\ Richer\altaffilmark{4},
	Gregory G.\ Fahlman\altaffilmark{5},
        Brad M. S. Hansen\altaffilmark{6},
        Jarrod Hurley\altaffilmark{7},
        Jasonjot S.\ Kalirai\altaffilmark{8,9},
        R.\ Michael Rich\altaffilmark{6},
        Peter B.\ Stetson\altaffilmark{5}}

\altaffiltext{2}{Space Telescope Science Institute, 
                 Baltimore MD; jayander@stsci.edu.}
\altaffiltext{3}{Department of Astronomy, University of Washington,
                 Seattle, WA 98195-1580.}
\altaffiltext{4}{Department of Physics and Astronomy, University of
                 British Columbia, Vancouver, BC, Canada.}
\altaffiltext{5}{Dominion Astrophysical Observatory.}
\altaffiltext{6}{Department of Physics \& Astronomy, 
  University of California at Los Angeles, Los Angeles, CA.}
\altaffiltext{7}{Centre for Astrophysics and Supercomputing, Swinburne
                 University of Technology.}
\altaffiltext{8}{University of California Observatories/Lick Observatory, 
                 University of California at Santa Cruz, Santa Cruz, CA.}
\altaffiltext{9}{Hubble Fellow.}

\begin{abstract}

We describe here the reduction methods that we developed to study the 
faintest red dwarfs and white dwarfs in an outer field of NGC6397, which 
was observed by \hst\ for 126 orbits in 2005.  The particular challenge 
of this data set is that the faintest stars are not readily visible in 
individual exposures, so special care must be taken to combine the 
information in all the exposures in order to identify and measure them.
Unfortunately, it is hard to find the faintest stars without also 
finding a large number of faint galaxies, so we developed
specialized tools to distinguish between the point-like stars and the 
barely resolved galaxies.  We found that artificial-star tests, while 
obviously necessary for completeness determination, can also play an 
important role in helping us optimize our finding and measuring algorithms.  
Although this paper focuses on this data set specifically, many of the 
techniques are new and might find application in other work, particularly 
when a large number of images is available for a single field.  

\end{abstract}

{\em Astrometry --- 
     globular clusters: individual (NGC 6397) --- 
     stars: low-mass ---
     techniques: image processing, photometric --- 
     white dwarfs }

\section{Introduction}
\label{SEC.INTRO}

NGC 6397 and M4 are by far the two closest globular clusters, and thus the
places where we can hope to study the faintest old stars.  In GO-5461 and
GO-8679, Richer et al.\ (1997) used the WFPC2 camera on board the {\it Hubble
Space Telescope} (\hst) to study the faint population of M4.  In 2005, our 
program GO-10424 used a similarly large number of orbits with a much more 
sensitive detector (the Advanced Camera's Wide Field Channel) to probe the 
faintest red dwarfs and white dwarfs in NGC 6397.  The program, and its initial 
results, are described by Richer et al.\ (2006).  

Here we present, in support of the detailed papers that have already been 
published (Richer et al.\ 2006, Hansen et al.\ 2007, Kalirai et al.\ 2007)
and others that will follow, the reduction procedures that we used for 
this massive data set.  The goals of this reduction are simple:\ find and 
measure the faintest stars in the field, and assess the completeness of 
our list.  Nonetheless, there are many complications that arise.  For 
instance, when searching for faint stars among stars that can be 20 magnitudes 
brighter, it can be very hard to separate the plentiful PSF artifacts 
from the relatively few faint stellar sources.  Also, at these faint 
magnitudes ($V \sim 29$), there is a large population of barely resolved 
background galaxies which must be distinguished from the true stars.

Our general approach here will be first to identify everything that could 
possibly be a faint star, then to subject this list to careful selection
criteria that leave a list of objects that seem to be truly stellar.  
Although our reduction here is specific to this particular data set, we 
believe that our procedures will also be useful for other large \hst\ data 
sets where the aim is to find the faintest objects possible.  Ground-based 
projects that have similar large data sets may also benefit from this approach.

The paper is organized along the lines of the reduction procedure itself.  
In Section \ref{SEC.OBS} we describe the observations, and in 
\S~\ref{SEC.PRELIM} we describe a number of preliminary reduction steps.  
Section \ref{SEC.FIND} describes the finding procedure, leading to a list 
of sources to be measured.  The measurement of magnitudes and positions is 
described in \S~\ref{SEC.PHOT}, and \S~\ref{SEC.WEED} describes several 
successive procedures for eliminating non-stellar objects.  Section 
\S~\ref{SEC.FINALLIST} covers the construction of the final stellar sample.  
In \S~\ref{SEC.COMPL} we use our measurement of artificial stars to 
estimate completeness.  Then \S~\ref{SEC.PMs} is devoted to the 
derivation of proper motions and their use.  A final section summarizes 
the reductions and the many scientific applications.

\section{Observations}
\label{SEC.OBS}

\hst\ program GO-10424 (P.I.\ Richer) was taken during 126 orbits between
13 March and 8 April 2005.  We observed a single field, 3\minspt17
E and 3\minspt93 S of the cluster center.  We wanted a field whose
density was high enough to give us a good number of white dwarfs, but
not so high that the background from the bright stars would prevent us
from finding the faintest stars.  This particular field has deep
archival observations from 1994 and 1997, with which we can measure
proper motions, and use them to remove field stars.

In order to go as faint as possible, we chose the wide-band filters
F606W and F814W.  Previous work had indicated that the faintest cluster
stars would be white dwarfs, so we balanced the exposure lengths so as
to be an optimum at the color of faint white dwarfs.  In each orbit we
took two long exposures in F814W and one in F606W.  Because the color of 
scattered Earth light is such that the F606W band is more vulnerable to it, 
the F606W exposure was always in the middle of the dark side of the orbit, 
when the sky is darkest.

There were 126 exposures in F606W, of length 616 to 769 seconds,
totaling 93,442 sec; in F814W there were 252 exposures, whose lengths
ranged from 616 to 804 sec, with a total of 179,704 sec.  The individual
exposures were dithered, by integral pixels plus fractions of a pixel.
We also took in each filter six short exposures, of lengths 1, 5, and 40
sec (two of each).  For each image the \hst\ pipeline provides a
bias-corrected and flat-fielded {\tt flt} image, along with a drizzled
{\tt drz} image (as explained in the {\it ACS Data Handbook}).  We
carried out our final measurements on the {\tt flt} images only, but
we used the first {\tt drz} image to provide a reference frame, and
the first pair of {\tt drz} images for photometric zero points.

\section{Preliminary Steps}
\label{SEC.PRELIM}

In this section we describe a number of procedures that we had to go
through before carrying out our full reduction:\ the mapping of warm
pixels, the construction of PSFs, the determination of the transformations 
from each exposure into the reference frame, the creation of image stacks, 
and the preparation of an additional set of images that have artificial 
stars added into them.

\subsection{Warm pixels}
\label{SUBSEC.WARM}

Because the accuracy at which we were aiming required that we correct
many more warm pixels than are flagged by the reduction pipeline, we
constructed and applied our own warm-pixel mask.  (Our dithering made this
possible, by causing a warm pixel to fall successively in different
parts of the scene.)  To identify warm pixels, we compared each pixel in
each exposure against the local background.  Any detector pixel that was
more than 4-sigmas bright (or dim) in at least half of the exposures was
flagged as bad.  Our warm-pixel list had 84,846 pixels, amounting to
0.5\% of the detector.  Since the number and characteristics of warm
pixels change with time, we are not including the list here, nor
describing it in detail, although it is available from the first author
by request.

\subsection{Construction of PSFs}
\label{SUBSEC.PSFs}

All of our measurements of star images were made by least-squares
fitting of a PSF.  For this we used an effective PSF, as originally
defined by Anderson \& King (2000, AK00), and generalized for the WFC
(Anderson \& King 2006, AK06).  Briefly, the effective PSF is a
continuous two-dimensional function, $\psi_E(\Delta x,\Delta y)$, which
tells us what fraction of a star's light will fall within a pixel that is
offset from the star's center by $(\Delta x,\Delta y)$.  We tabulate our
effective PSF model on a grid in $(\Delta x,\Delta y)$ space, and evaluate 
the PSF in between these points by bi-cubic interpolation.

A complication of the WFC PSF is that it varies with position on the
detector.  A small part of the variation is caused by optical
aberrations, but most of it comes from charge diffusion, whose amount
depends on the local thickness of the detector (Krist 2003).  We found
in AK06 that we could model the WFC PSF quite well with an array of
9$\times$5 fiducial PSFs over each of the two 4096$\times$2048-pixel 
chips.  We used the procedures in AK00 to construct such a time-average 
PSF for each filter, using the 252 dithered F814W images and the 
126 dithered F606W images.  

In addition to its spatial variation, we found in AK06 (which was
based partly on these data) that the PSF also varies by a small amount 
from exposure to exposure, most likely due to thermal breathing or pointing 
instabilities.  This temporal variation of the PSF is typically such 
that the fraction of light in the central pixel of a star which is centered 
on a pixel can range from 18\% to 22\% in F814W, and by similar amounts 
in F606W.  Fortunately, the variation across the chip seems to be 
largely constant with time, so that one can construct for each individual 
exposure a single ``perturbation'' PSF that can be added to the spatially 
variable PSF described above, to provide more accurate PSFs for that image.

\subsection{Transformations between images}
\label{SUBSEC.TRANS}

There is a fundamental difference between bright stars and faint stars.
Bright stars can be easily identified in almost every exposure, whereas 
in some images faint stars may not stand out above the noise.  
To find and measure the faint stars, we will need to analyze all the 
exposures together.  We therefore need a way to transform from positions 
in each exposure into the adopted reference frame, and vice versa. 
So in this section we will first analyze the relatively bright stars to 
establish the astrometric and photometric transformations from each 
exposure into a reference frame. 
 
We chose the first drizzled image ({\tt j97101bbq\_drz.fits}) as the 
astrometric basis for the reference frame, since {\tt drz} images have
been corrected for distortion.  Its image header also allows 
us to map pixel coordinates into RA and Dec.  Our PSF model is appropriate 
for {\tt flt} images, but not for {\tt drz} images, so we measured 
simple centroid positions for a few hundred bright but unsaturated stars 
in this image.  These positions will serve to specify the orientation,
scale and zero point of the reference frame.

Next, we needed to measure positions for the bright stars in each exposure,
so that we could relate each exposure to the reference frame.  For this,
we use the PSFs described above and the FORTRAN program img2xym\_WFC.09x10
(which is described in AK06) to measure positions and fluxes for the sources
with more than 100 counts in each exposure.  We then found the common stars 
between each image list and the reference list above, and used their 
coordinates in the two systems to define a general 6-parameter linear 
transformation from the distortion-corrected frame of each exposure into 
the reference frame.  

The centroid positions that originally defined the reference frame were 
not very accurate.  So, in order to replace them with more accurate
measurements, we used the above transformations to collate the 
individual star lists in the reference frame, identifying a star wherever 
180 out of 252 F814W lists found a source.  We found 7,840 such bright
stars and determined for each an average position in the reference frame.
These positions are much more accurate (in a relative sense) than the 
centroid positions from the {\tt drz} image, so we adopted these 
average positions as the new reference frame.  We then constructed new 
transformations from each exposure into the improved reference frame.  With 
these new transformations, we again found improved average reference-frame
positions for each star.  The RMS residuals about these average positions 
were now less than 0.01 pixel in each coordinate, and the photometric 
residuals were about 0.01 magnitude for the bright stars (S/N$>$100).  These 
small residuals confirm that we now can transform transparently from the 
individual exposures into the reference frame, and vice versa.   

Now that we had an accurate reference frame, we repeated the above 
bright-star finding for the F606W exposures, and constructed average F606W 
magnitudes for the stars as well.  The instrumental photometric zero points 
were set to correspond to the first exposure of each filter.  Since the 
photometry routine automatically measures saturated stars, we include the 
saturated stars in this list, but we keep in mind that only the stars 
fainter than $-14$ (instrumental magnitudes, $-2.5\log(DN)$) are reliably 
measured.  (For most of this paper, we will keep our photometry in the 
instrumental system, since it is a more natural system for evaluating 
photometric quality in terms of signal-to-noise.  We will calibrate the 
phtometry in \S~\ref{SSEC.ZP}.  For convenience, we provide the calibrated 
scale on many of the plots.)

\begin{figure}[!t]
\epsscale{1.00}
\plotone{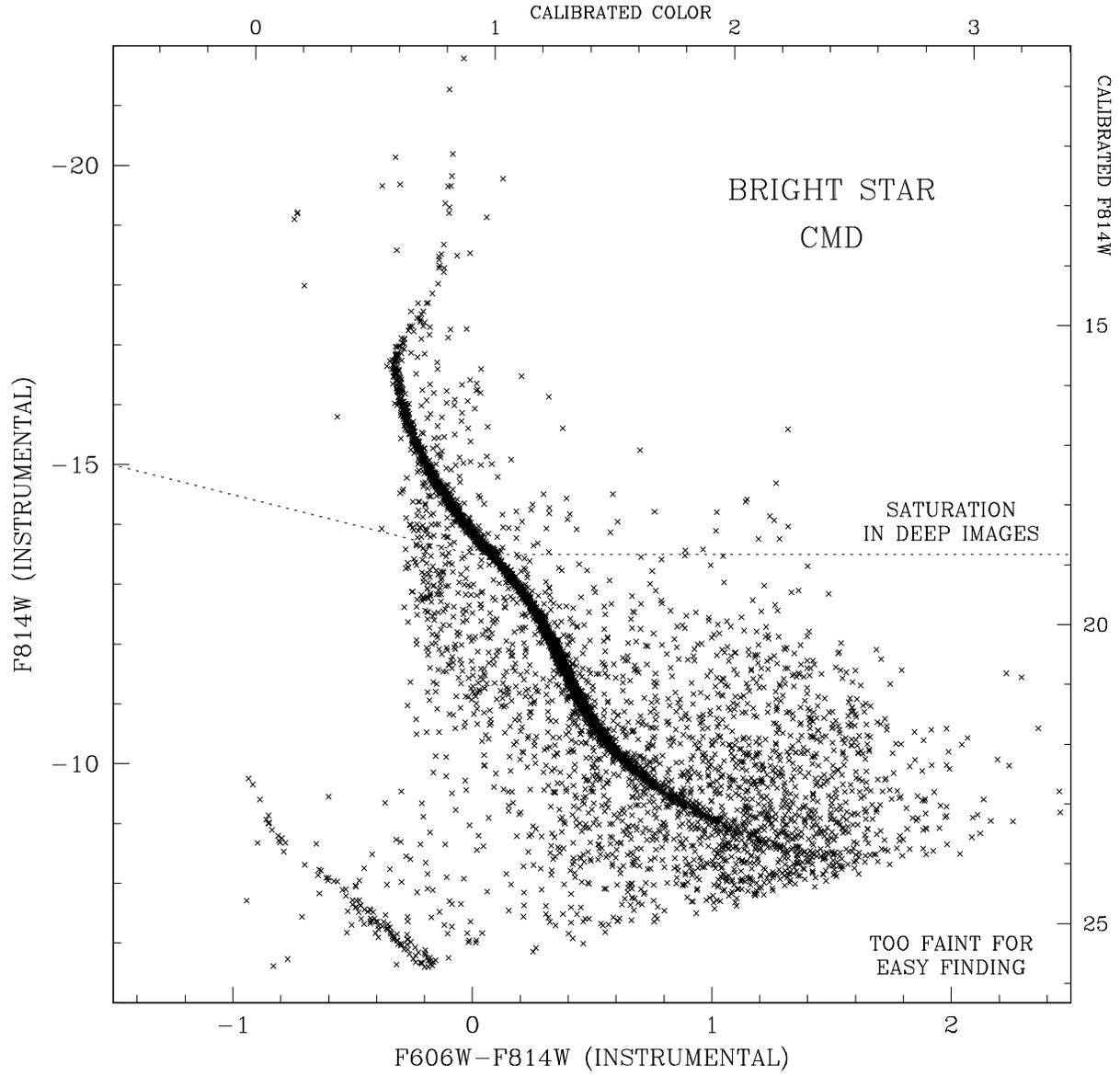}
\figcaption{A CMD for the bright stars that are easily found in
            every exposure.  The short exposures have been used to measure
            fluxes for the stars that were saturated in the deep images.
\label{FIG.BRIGHT}}
\end{figure}

Figure~\ref{FIG.BRIGHT} shows the CMD for the stars found this way.  
The photometry and astrometry in this list are essentially the best that 
can be done for the brighter stars, since we can fit for both their
position and flux in each individual exposure.  The rest of this paper 
will focus on how best to measure the fainter stars, making use of the 
transformations established by the brighter stars.

\subsection{Image stacks}
\label{SUBSSEC.STACKS}

Although all our final measurements were made on the individual exposures,
it was useful to construct stacks so that we could examine the images of 
stars and galaxies that were hard to see clearly in individual images.
The stacks were invaluable in helping us discriminate between objects
the computer should classify as a star, and objects that should be rejected 
as a galaxy or PSF artifact.

Since the raw {\tt flt} images were undersampled, we decided to
supersample the stack by a factor of two in each coordinate.  (To avoid 
confusion between pixels of a stack and pixels of an image, we will 
refer to the pixels of the stack as subpixels.)  The average number of 
image pixels contributing to the evaluation of each subpixel of the
F814W stack was 252/4 = 63; for the F606W stack the number was of course
half as large.  To cover the reference frame fully, the stack was
8500$\times$8500 subpixels.

There is no unique way to combine many exposures into one stacked image.
Drizzle (Fruchter \& Hook 2002) provides one approach, Lauer (1999)
another.  Different approaches preserve different properties (e.g.,
sampling, flux, etc.).  Since we use the stack in a qualitative rather
than a quantitative way, the details of how we constructed the stack are
not terribly important.  Briefly, we used the transformations found in
the previous section to determine where each pixel in each exposure
mapped into the stack frame.  The photometric component of the
transformations scaled each exposure to match the first exposure.  As an
initial estimate of the stack, we took the $\sim63$ F814W and $\sim31$
F606W pixels that mapped into each subpixel and assigned that subpixel
the sigma-clipped average value.  This is akin to using drizzle with
$\tt pixfrac$=0.

This initial guess could suffer from sampling biases if the dither does
not cover each pixel evenly.  So we iterated the procedure once.  We
used the transformations to resample the stack into the frame of each
exposure.  The difference between the actual exposure and this simulated
image is an indication of how this image would influence the stack to
change.  We then adjusted each subpixel in the stack by the
sigma-clipped average of the F814W and F606W residuals that mapped into
it.

These stacks served as approximate mean images that made it easy to
examine any region in the field.  Even though the stack-construction 
procedure was flux-conserving and the final resolution is nearly 
Nyquist sampled, we found that photometry on the stacks was not as 
accurate as simultaneous photometry using all the individual images together.  
This is probably due to the fact that we have an exquisitely accurate
PSF model for the {\tt flt} images, but no appropriate PSF for the
stack (such a PSF would be complicated by the resampling process
inherent in the creation of the stack).

\begin{figure}[!t]
\epsscale{1.00}
\plotone{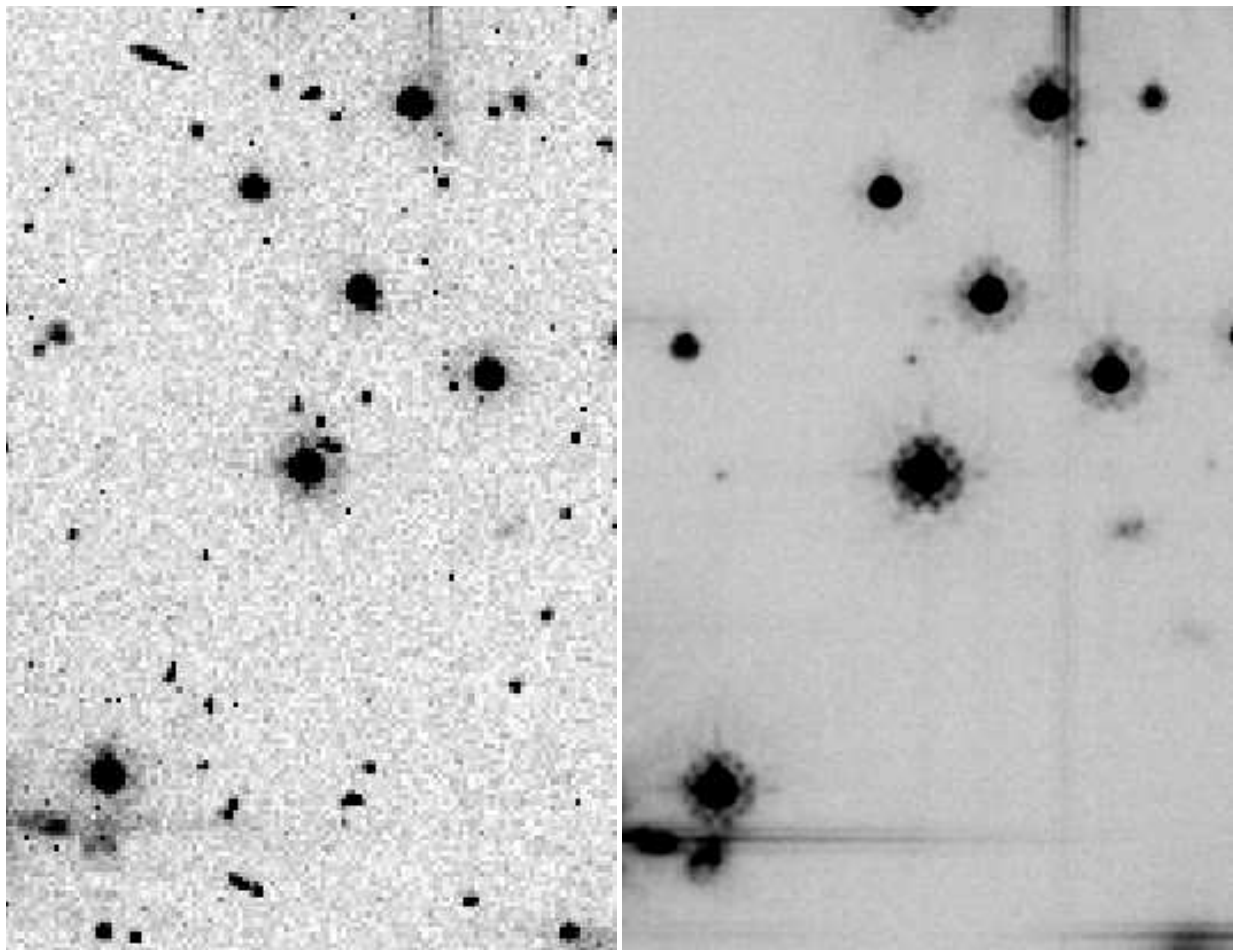}
 \figcaption{({\it  left}) A small region in the first F814W exposure of 
                           the series. 
             ({\it right}) The same region in the stacked image.
\label{FIG.STACK.COMP}}
\end{figure}

Figure~\ref{FIG.STACK.COMP} compares a portion of the stack with a single 
exposure.  It is clear from this image that there are many non-stellar 
features in the field, such as galaxies, PSF artifacts, and diffraction 
spikes, which an automated searching routine could easily confuse with 
stars.

\subsection{Artificial stars}
\label{SUBSSEC.AS}

One tends to think of the addition of artificial stars (ASs) primarily in
conjunction with completeness tests.  We did indeed use AS tests for that 
purpose, but they also played a crucial role in helping us develop our 
finding and measuring strategy.  In addition, the large number of AS tests
we performed allowed us to model the distribution of observed stars in the 
WD region in terms of the input distribution (see Hansen et al.\ 2007).

We wanted to insert as many artificial stars as possible in a single run, 
so as to minimize the number of runs necessary.  The main considerations
in performing AS tests are (1) that they be treated in an identical manner 
to the real stars and (2) that they not interfere with each other, thereby
affecting the crowding conditions they are intended to measure.  So, we
constructed a grid of positions in the reference frame, with a
separation between nearby stars of 10 pixels in each direction.  Since
the faint stars we are considering affect only a handful of pixels,
there is no way they will interfere with each other.  We inserted the
stars with a flat luminosity function in F814W between magnitudes $-11$
and $0$, and with an F606W magnitude that placed the star on the white
dwarf cooling sequence (WDCS).  As it turned out, we used only the F814W 
images for finding and for image discrimination, so that the choice of 
AS colors did not not affect our completeness estimates. 

We tilted the grid by 5 degrees, in order to avoid alignment with rows and 
columns of the detector.  Finally, to ensure that the artificial stars were 
reduced the same way as the real stars we added each artificial star to 
each of the 252 + 126 {\tt flt} individual exposures as follows:  We
transformed the location and flux of each artificial star into each exposure, 
and added a scaled version of the PSF with the appropriate noise.  The total 
number of artificial stars in the pattern was 176,644.

\begin{figure}[!t]
\epsscale{1.00}
\plotone{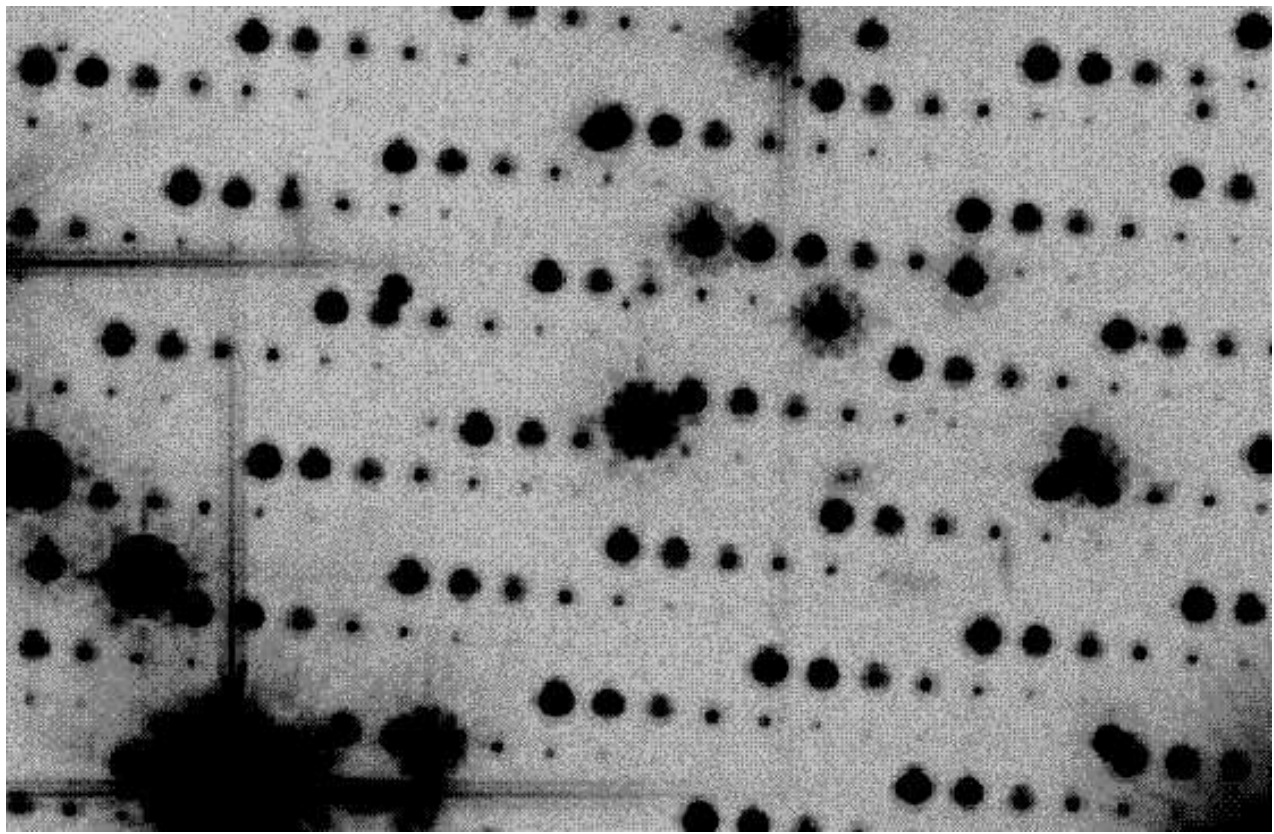}
\figcaption{The stack of the F814W images that have artificial stars
added.  The first star in each set of 11 has an F814W magnitude between
$-11$ and $-10$, the second star between $-10$ and $-9$, etc., making it
easy to assess by eye at what magnitude stars become hard to find.
\label{FIG.ARTSTACK}}
\end{figure}

We also made a stack image from the AS images in each filter.  We
were thus able to carry out on the AS images every procedure that we
used on the real-star images.  Figure~\ref{FIG.ARTSTACK} shows a portion
of one of the stacks of AS images.

\section{Finding the Stars}
\label{SEC.FIND}

The goal of this reduction is to arrive at the most complete possible
list of positions and fluxes for stars in the field, along with an estimate 
of the incompleteness of the sample.  The need to assess incompleteness means
that we must have an automated finding algorithm which can be applied to
the artificial stars in a fashion identical to the way we handle the
real stars.  This automated finding algorithm must be able to
identify the faintest possible stars, while at the same time ensuring
that non-stellar objects will not find their way into the sample.

We therefore divide the list-construction task into three steps:  finding,
measuring, and weeding.  Our first step is to go through the data set and 
find all features that could possibly be stars.  We then measure positions, 
fluxes, and other properties for each of these objects.  Finally, in the 
weeding stage, we examine the surroundings and shape of each in order to 
remove those that are likely not to be genuine stars.  The present section
covers the initial finding.

\subsection{The peak map}
\label{SSEC.PEAKMAP}

The first task is to find all of the features that could possibly be stars.
At this point, it is worthwhile to consider what effect a barely detectable
star will have on this data set.  Since the WFC images are slightly
undersampled, a large fraction of each star's flux will be concentrated
in its centermost pixel.  In a single exposure, a faint star may or may not
be able to push its central pixel above the noise.  If the star is so faint
as almost never to do so, then there is no way we will detect it.
However, if it can a generate a local maximum in a statistically significant
number of images, then we have some chance of detecting it.

We therefore begin by defining a local maximum as any pixel that is higher
than any of its eight neighbors.  Such pixels are very numerous --- on
average, one pixel in nine, in any region that is mostly background
noise --- but if a local maximum occurs in the same place in many
exposures, it is an indication of the presence of a faint star.  The
challenge in finding the faintest stars, then, is to search for places
in the field where there are significantly more local maxima than we
would expect from random noise alone.  As we have said, with a low
detection threshold we will catalog a large number of spurious peaks,
but we are willing to accept them and then weed out the spurious ones,
thereby reaching deep into the noise to detect stars that we can find
only because we have so many exposures of the same region.

To identify these candidate stars, we construct what we call a ``peak map.''
This is a map of how often a local maximum occurred at a particular
place in the field.  To construct this map, we go through each of the
252 F814W exposures pixel-by-pixel.  Each time we find a pixel to be a
local maximum, we use the transformations to calculate the corresponding
location in the reference frame, and add one count to the closest
peak-map pixel.  We sampled the peak map in the same way as the stack,
with pixels that are half the linear size of the image pixels so that
the peak-map pixelization would not affect our resolution.  (As an
example of how we used our list of warm pixels, we did not allow any of
them to contribute to this peak count.)

A faint star does not make a local peak in every image, and random peaks
add a background level of $252/(9\times4)=7$ (the division by 4 because
of the supersampling); but even with this background, faint stars 
stand out in a statistically significant way.  A non-stellar object such
as a galaxy with a sharp center is also likely to show up in the peak map; 
we admit it at this stage but count on eliminating it later.

\begin{figure}[!t]
\epsscale{1.00}
\plotone{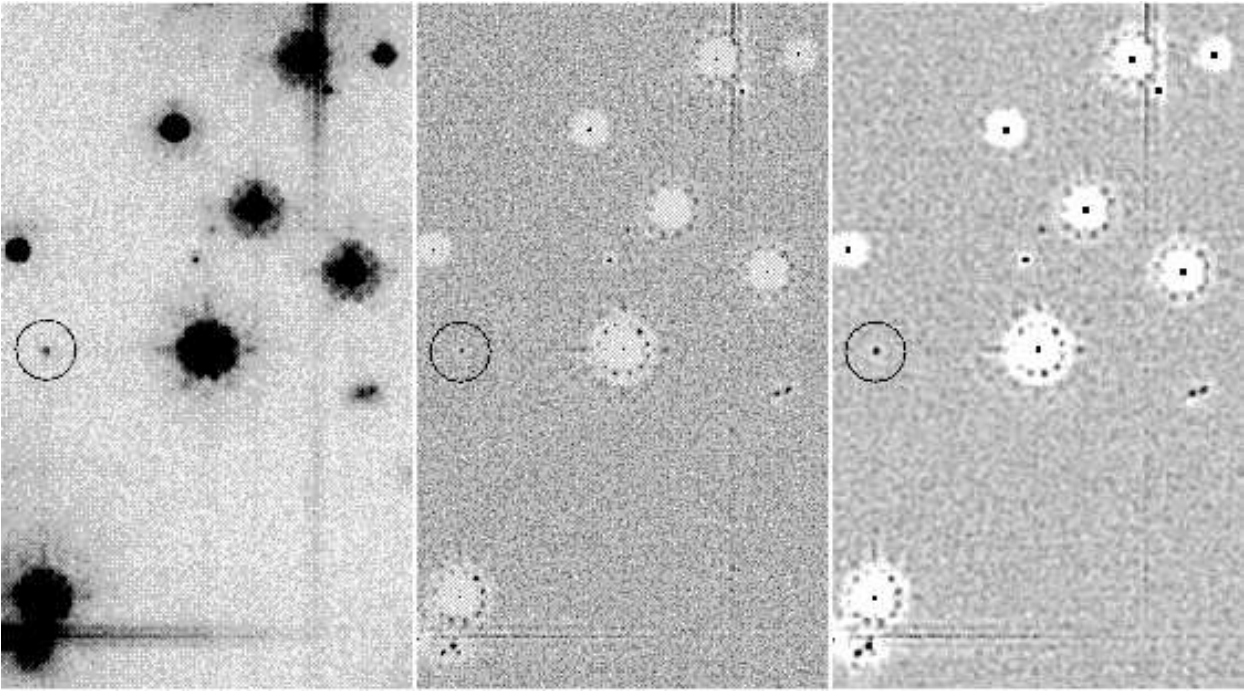}
\figcaption{({\it left}) A portion of the stack centered on (3512,1577). 
             ({\it middle}) The raw peak map for the same region.
             ({\it right}) The 3$\times$3-overbinned peak map.  
             The faint star on the left (circled) is a WD near 
             the very bottom of the sequence.  It generated 
             a peak in 180 out of 252 exposures.
\label{FIG.PKMAP}}
\end{figure}

The middle panel of Figure~\ref{FIG.PKMAP} shows this raw peak map, 
alongside the F814W stack.

%
%

\subsection{Analyzing the peak map}
At this point we needed a way of analyzing the peak map that would allow
us to collect nearly all of the genuine stars, but very few of the random
or artifact-related peaks.  We explored many ways of extracting a source
list from the peak map.  In particular, the peak map that was made from the 
images that contained artificial stars proved particularly useful for
telling us  what signature stars of different brightness would have on the 
peak map.  

We found that by overbinning the peak map by 3$\times$3 (in other words,
just like 3$\times$3 boxcar smoothing, but without dividing the sum by
9) we were able to optimally highlight the signal from the faintest
stars.  This overbinning was necessary because stars do not always fall
at the centers of pixels, either in the individual exposures or in the
peak map, and we want to consider all of the peaks that each star
generates in order to maximize its signal.

We adopted a two-parameter finding algorithm.  In order to be included in
our initial list, a source had to (1) generate a peak in at least 90 out of
252 F814W images, and (2) be at least 15 subpixels (7.5 WFC pixels) away 
from any more significant source.  These parameters represent a balance 
between including as many genuine stars as possible, without including too 
many non-stellar features.  From the artificial-star tests, we found that 
many stars with about 90 peaks were recovered with positions and fluxes
that were close to the input values.  With a background of 63 in the 
overbinned map, this represents a 3.4-$\sigma$ detection, so lowering 
the threshold below this would dramatically increase the noise contribution 
to our sample.  The second condition naturally rejects the PSF artifacts 
found around the bright stars.  Since the typical separation between stars 
in this field is about 25 pixels, this excludes many non-stellar objects
without excluding many of the detectable stars.

The above strategy resulted in a list of 48,785 sources.  For each source, 
we have a position in the reference frame, based on the centroid of the 
found peaks.  The artificial-star tests indicate that this position is good 
to about 0.1 pixel even for the faintest stars.   All of these peaks 
went forward to the photometry stage; 8,399 of them would survive the 
later weeding-out processes that we will describe in \S~\ref{SEC.WEED}.

Finally, we note that we explored many different ways of constructing
the above source list.  We first thought to derive it from the stack image, 
but the peak-map approach seemed more ``democratic'' in its selection, 
in that the individual-exposure peaks that are higher above the background
(perhaps due to just-undetected CRs) would have a disproportionate
influence on a stack, but not on the peak map, where each image gets one
vote.  Also, we initially included the F606W images in the peak-map 
analysis, but found from the artificial stars that using the F606W images 
did not add a significant number of stars.


\section{Photometry}
\label{SEC.PHOT}

We initially tried doing photometry on the stacked images, but the 
resulting color-magnitude diagram indicated that the faint stars were 
nearly lost among objects that were not real.  We therefore made all 
our final measurements on the individual {\tt flt} images.

To measure stars that were too faint to see in a single exposure, we
collected all of the information for a given star, in each of the two
filter bands, as follows.  We used the positional transformations and
the distortion corrections to calculate the position of the star in each
individual exposure, and extracted a 7$\times$7 array of pixels around
that center.  Since exposure times differed slightly, we normalized
each pixel array to the same exposure time.  Of the 49 pixels from
each exposure, we used the sigma-clipped average of the 28 pixels between
$r=1.5$ and $r=3.5$ pixels of the central pixel as the sky level, and 
subtracted this value from each of the innermost 9 pixel values.  
These $252\times9$ pixel values (for F814W, or $126\times9$ for F606W) 
were the data that we used to find the flux of the star.

As we indicated earlier, our effective PSF tells us the fraction of a 
star's light that is expected to fall in a pixel centered at an offset 
($\Delta x$, $\Delta y$) from its center.  The flux in each pixel in
each exposure $n$ should be described by:
$$  
P_{ij,n} = f_* \psi_{ij,n} + s_n, 
$$
where $f_*$ is the star's flux, $s_n$ is the sky value (found above), and
$\psi_{ij}$ is the fraction of light that should fall in that pixel.  This 
is the equation for a straight line with a slope of $f_*$ and an intercept 
of $s_n$.  

\begin{figure}[!t]
\epsscale{1.00}
\plotone{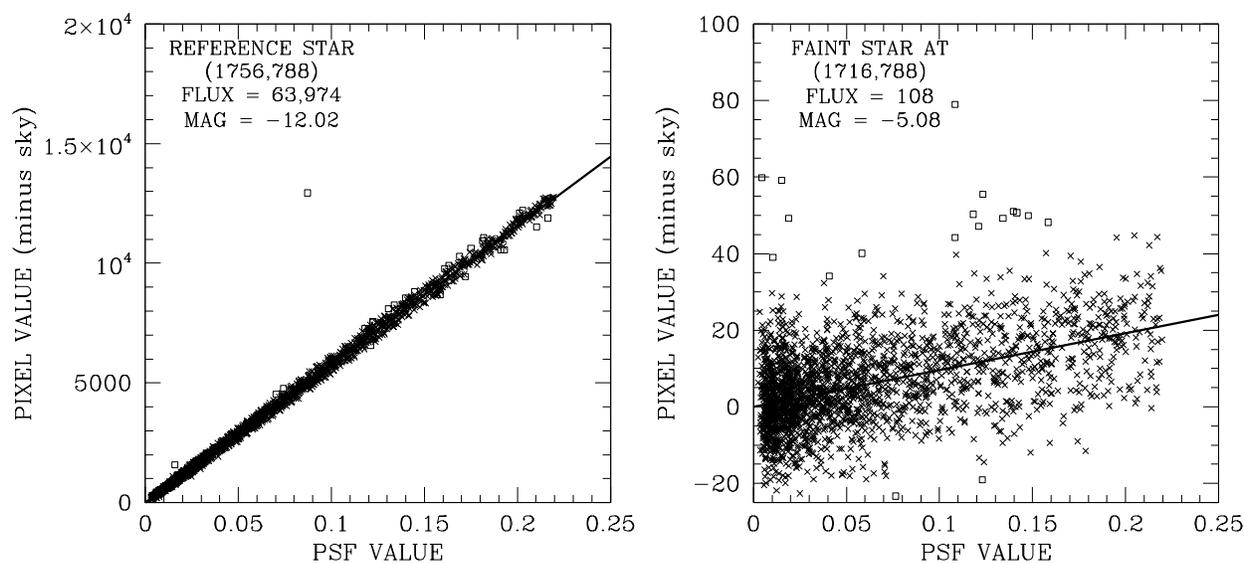}
\figcaption{({\it left}) A plot of observed pixel value against PSF value
                   ($\psi_{ij}$) for the bright star at the center 
                   of the field shown in the previous figures.  The flux
                   is the slope of this relation.  The open squares are
                   the points that were rejected from the fit.
            ({\it right}) Same for the faint WD.
\label{FIG.STARFIT}}
\end{figure}

Figure~\ref{FIG.STARFIT} shows how the set of pixels for a star is fit
by the PSF model.  The tight linear relation for the bright star on the
left shows that our PSF model is extremely accurate and also that our
astrometric and photometric transformations are able to properly relate
each pixel in each exposure to the model.  This demonstrably good fit for
the bright stars means that we can trust our model to measure accurate
fluxes for the fainter stars.  The star on the right is one of the
faintest white dwarfs in the field.

We fit the flux $f_*$ for each star (the slope in Figure~\ref{FIG.STARFIT})
by a least-squares fit to all the pixels, taking into account the expected 
noise in each pixel.  We iteratively rejected the points that are more than 
4-$\sigma$ discordant with the best-fitting model.  In addition to fitting
for the flux, we also look at residuals from the fit to construct some
shape parameters that we will use to differentiate stars from galaxies
in \S~\ref{SEC.WEED}.

\begin{figure}[!t]
\epsscale{1.00}
\plotone{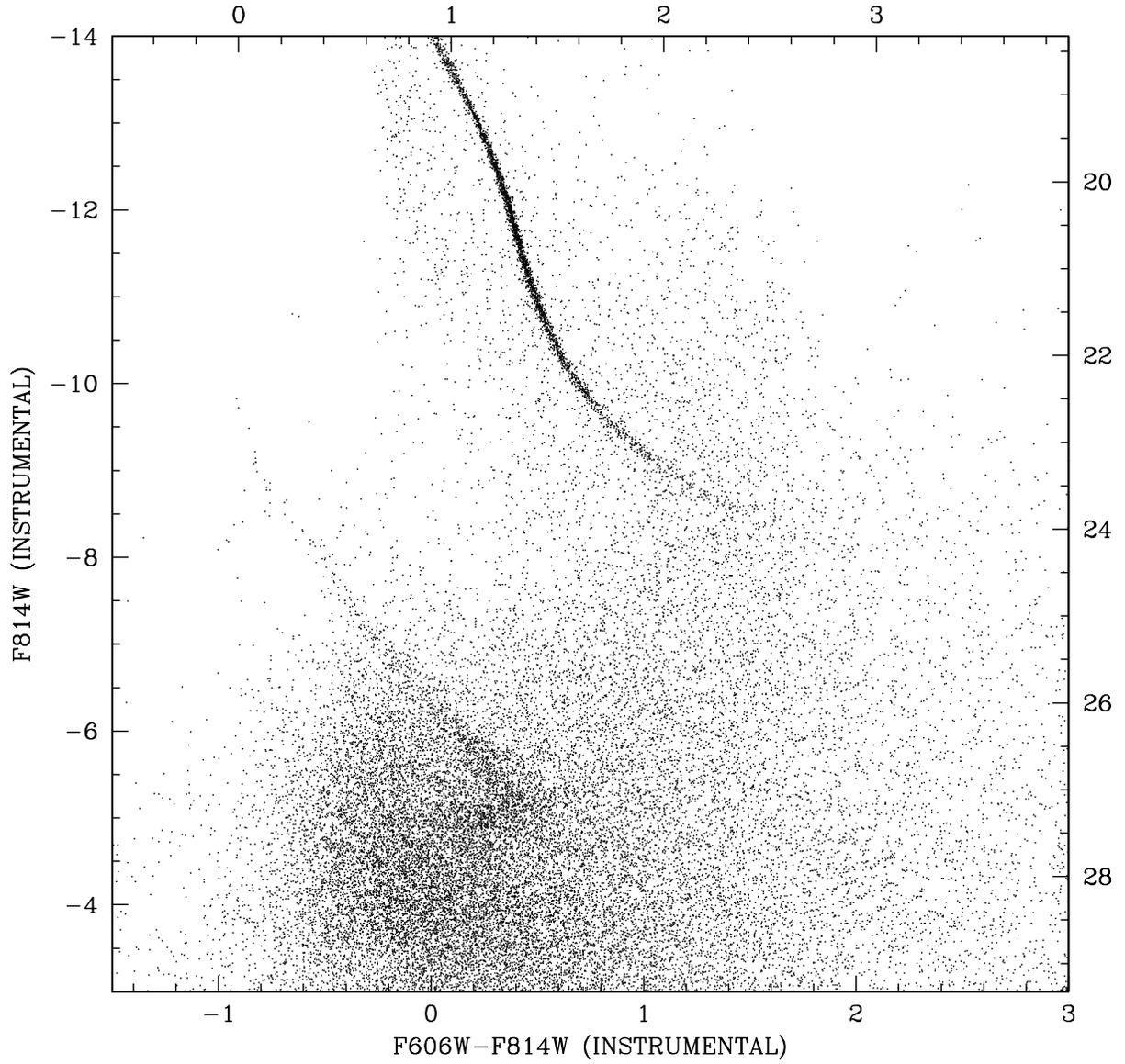}
\figcaption{The CMD of all the sources found in our peak-based finding 
            scheme, with fluxes measured by fitting the PSF simultaneously 
            to all images.  (The calibrated F814W photometry scale is 
            shown on the right.)
\label{FIG.RAWCMD}}
\end{figure}

Figure~\ref{FIG.RAWCMD} shows the raw CMD for all 48,795 of our sources.  
The WD sequence shows up distinctly, but there are many cluttering sources 
as well.  The following section will show how we winnowed out the non-stellar 
objects.


\section{Weeding Out Spurious Objects}
\label{SEC.WEED}

In order to include the faintest stars in our list, we had to cast our 
net so wide that we necessarily included many objects that are not 
stars:\ galaxies, PSF artifacts, features on diffraction spikes, etc.  
As a first step toward eliminating these, we used the stacked images to
examine many of the 48,785 objects that our procedures had found.
First, we saw that our list did include almost all of the stars that
could have been found by eye, except for those with very close neighbors.
But in addition, our selection criteria had admitted a large number of 
artifacts around bright stars, and also many bumps along diffraction 
spikes.  Finally, many of the clearly real objects were galaxies rather than 
stars.  On the basis of this examination, we developed three tests to 
separate true stars from artifacts and galaxies.  To eliminate artifacts 
we used the bright-star-proximity and the spike test; we eliminated 
galaxies with the shape test.

\subsection{The bright-star-proximity test}
\label{SSEC.PROX_WEED}

Faint detections found close to bright ones pose two kinds of problem:\ 
(a) many of them are not stars but instead are artifacts in the outer 
part of the PSF of the bright star, and (b) a faint star that is too close 
to a bright one cannot possibly be measured well, and should be discarded 
for this reason.

\begin{figure}[!t]
\epsscale{1.00}
\plotone{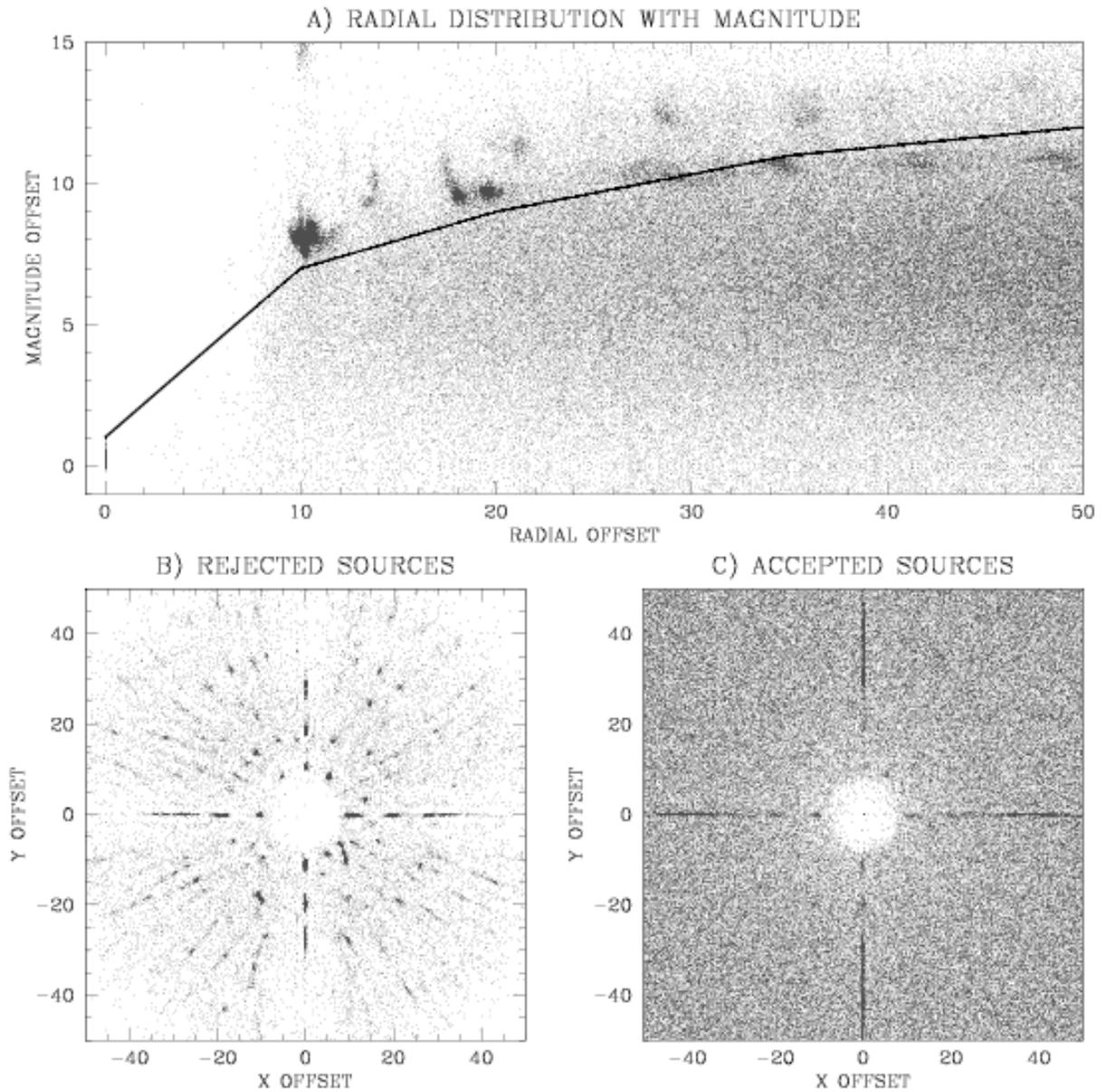}
\figcaption{({\it top}) The distribution of position and magnitude offsets for 
                  sources found around bright stars.   Detections above 
                  the line are too close to brighter stars to be reliable.
         ({\it bottom left}) The spatial distribution of rejected detections.
         ({\it bottom right}) The spatial distribution of accepted detections. 
                  See \S~\ref{SSEC.PROX_WEED} for a discussion.
\label{FIG.PROXWEED}}
\end{figure}

To address these problems, we took the stars that were brighter than
F814W magnitude $-10$ (10,000 $e^-$) and tabulated all the fainter objects 
in our list that were within 50 pixels of one of these ``bright'' stars.  
When the magnitude differences are plotted against separation, in the 
upper part of Figure~\ref{FIG.PROXWEED}, many distinct clumps show up, 
indicating that artifacts tend to occur at specific combinations of 
separation and magnitude difference.  The upper set of clumps correspond
to radial features in the PSF, and the lower set to features along 
diffraction spikes.  We drew the separator line in such a way as to exclude 
the radial features.  Removing things along the diffraction spikes with 
a purely radial discriminating function would be unnecessarily harsh and
would disqualify many valid, measurable sources.  We will remove the 
features along the spikes in the next section, with a more targeted 
procedure.  The bottom-right plot shows that this discrimination removes 
all PSF artifacts, except those along the diffraction spikes.  Of the 
48,785 objects with which we started, 27,222 survived this first weeding 
stage.

\subsection{The spike test}
\label{SSEC.SPIKE_WEED}

A generic property of the artifacts that lie on the diffraction spikes
of bright stars is that they are elongated along the direction of rows
or columns, which are the directions in which the spikes run.  We
therefore devised a statistic that measures the tendency of the light
around an object to be extended along the rows and columns.

\begin{figure}[!t]
\epsscale{1.00}
\plotone{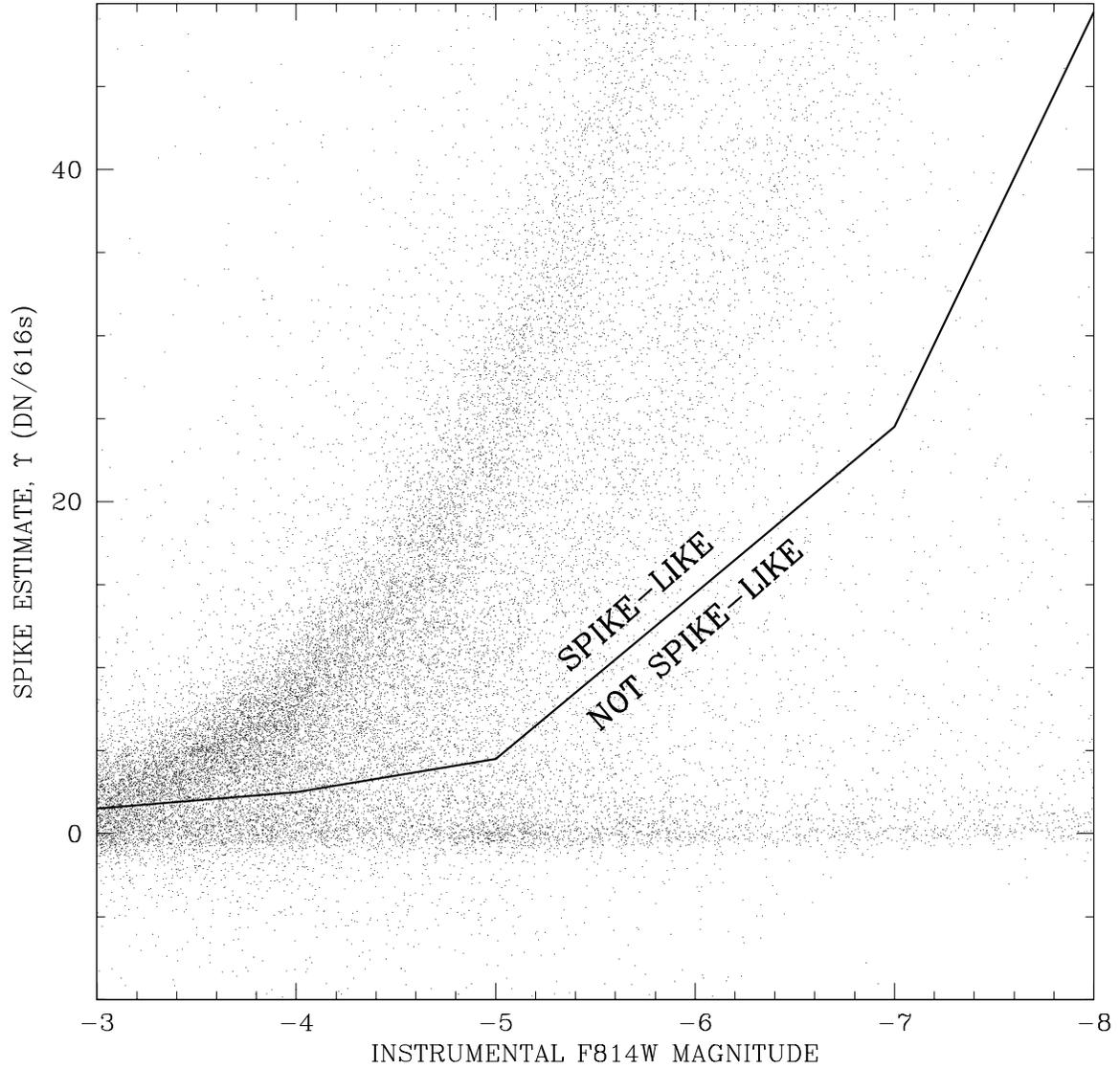}
\figcaption{The distribution of the spikiness index $\Upsilon$ as a
            function of fit instrumental magnitude for all the detected 
            sources.  See \S~\ref{SSEC.SPIKE_WEED} for a discussion.
\label{FIG.SPIKEWEED}}
\end{figure}

For this we took the F814W stacked image and binned it 2 by 2 to get a
$4250\times4250$ image.  For each of the remaining 27,222 objects, we
analyzed the 21$\times$21 surrounding pixels to determine the likelihood 
that it is on a row-like or column-like diffraction spike.  To estimate 
the row-oriented spike contribution, we found the sigma-clipped average 
value for the 18 pixels with $|\Delta i| \ge 3$ and $\Delta j = 0$
(where $i$ is the column that the pixel is in and $j$ is its row), 
and subtracted the sigma-clipped average value of the 324 background pixels 
($|\Delta i| \ge 3$ and $|\Delta j| \ge 3$).  We computed a similar 
column-oriented spike contribution, and added the two to get a combined 
spike estimate.  When these estimates are plotted against magnitude, the 
spike-like objects stand out well, except at the very faint end.
We drew the separator line shown in Figure~\ref{FIG.SPIKEWEED}
in an effort to include everything that was consistent with the bulk 
of the stellar distribution.  We inspected the remaining sources in the 
images and found that this did an excellent job removing features on 
spikes.  This step eliminated 11,773 of the remaining 27,222 objects.

\subsection{Shape tests}
\label{SSEC.SHAPE_WEED}

The spike and bright-star-proximity tests thus reduced our list of
objects to 15,449.  Examination of the stacks showed that some of these
were clearly galaxies.  The image of a galaxy differs from that of 
a star in two ways that are easily tested:\ it has more light in its 
envelope, relative to the core, than does the PSF of a star, and it 
may be elongated.  We therefore devised criteria of sharpness and 
roundness; to avoid confusion with the SHARP and ROUND parameters 
that DAOPHOT produces (Stetson 1987), we called ours CENXS and ELONG.

As the first step in calculating the value of the sharpness parameter
CENXS for a star, we located in each of the images all the pixels whose
centers were within 0.5 pixel of the calculated center of the star in
that image, and subtracted from each of them the value of the fitted
PSF.  We then took a sigma-clipped mean of these residuals; CENXS is 
this mean divided by the total flux of the star.  It is zero for an 
object that agrees perfectly with the PSF, and is negative for an 
extended object.

To calculate our roundness parameter ELONG, we located in each image
those pixels, on opposite sides of the center of the star, whose centers
fell within a pair of pixel-size squares centered 1 pixel away, and
again took a sigma-clipped mean of these pixels over all of the images.
We did this at position angles 0, 45, 90, and 135 degrees; the value of
ELONG is the largest of the four, less their mean, divided by the total 
flux of the star.  For an object that is round, it is close to zero, 
while its value increases with the elongation of the object.

\begin{figure}[!t]
\epsscale{1.00}
\plotone{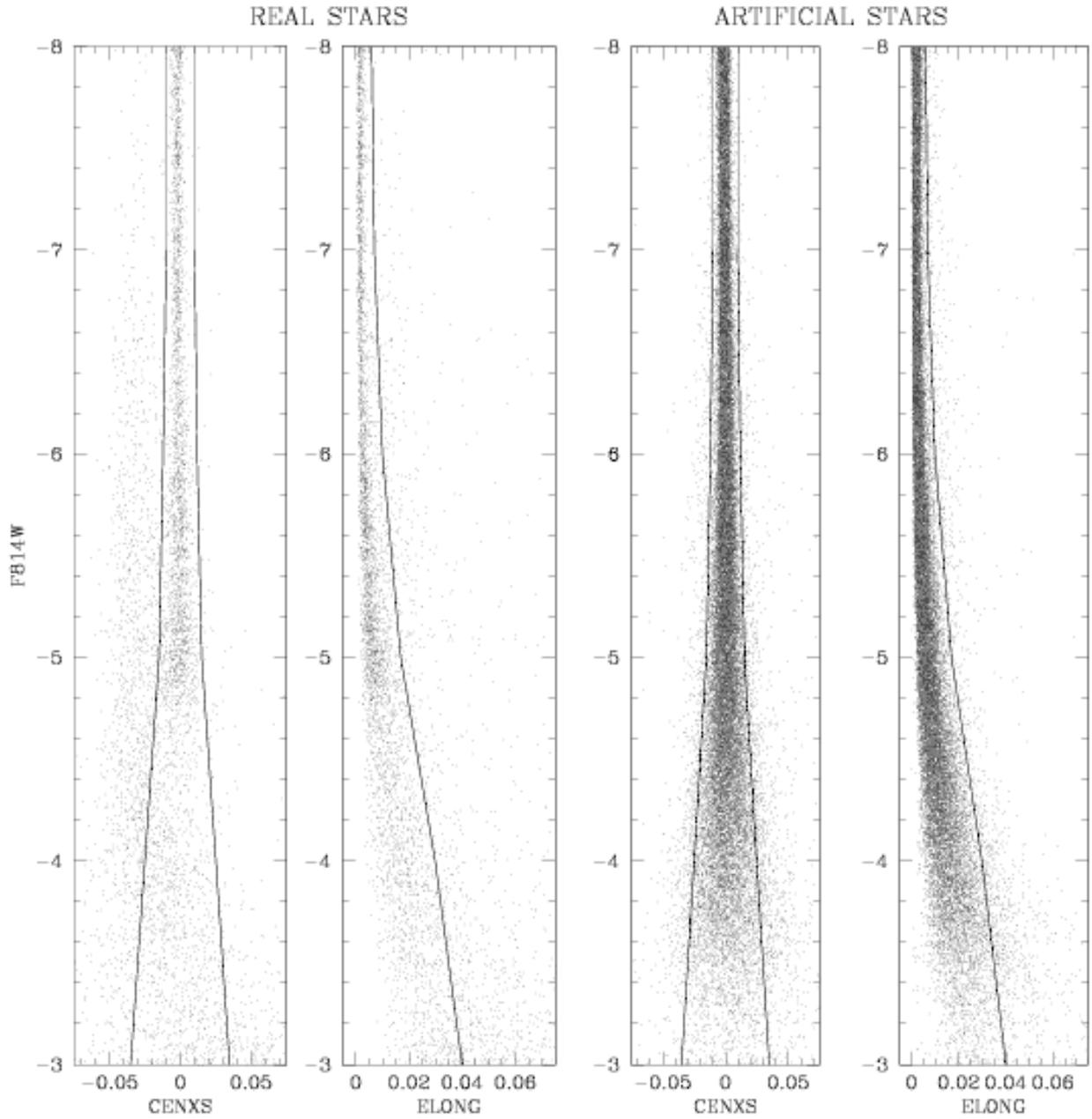}
\figcaption{The shape diagnostics for the real and artificial stars.
            The real stars are shown as a function of their observed F814W
            magnitude, the artificial stars are shown against their 
            inserted F814W magnitude.
\label{FIG.SHAPEWEED}}
\end{figure}

Figure~\ref{FIG.SHAPEWEED} shows the distributions of CENXS and ELONG
for the objects that survived up to this stage, along with their
distributions for the artificial stars.  In the artificial-star diagrams,
we drew in subjectively chosen lines that would include nearly the entire
point-source distribution, and we used these lines as criteria that would
discriminate between stars and galaxies, among the real objects.  Of the 
15,449 objects that entered this stage, 8,768 survived this final test.  
(In this test, CENXS was very much the more effective of the two criteria.)

\begin{figure}[!t]
\epsscale{1.00}
\plotone{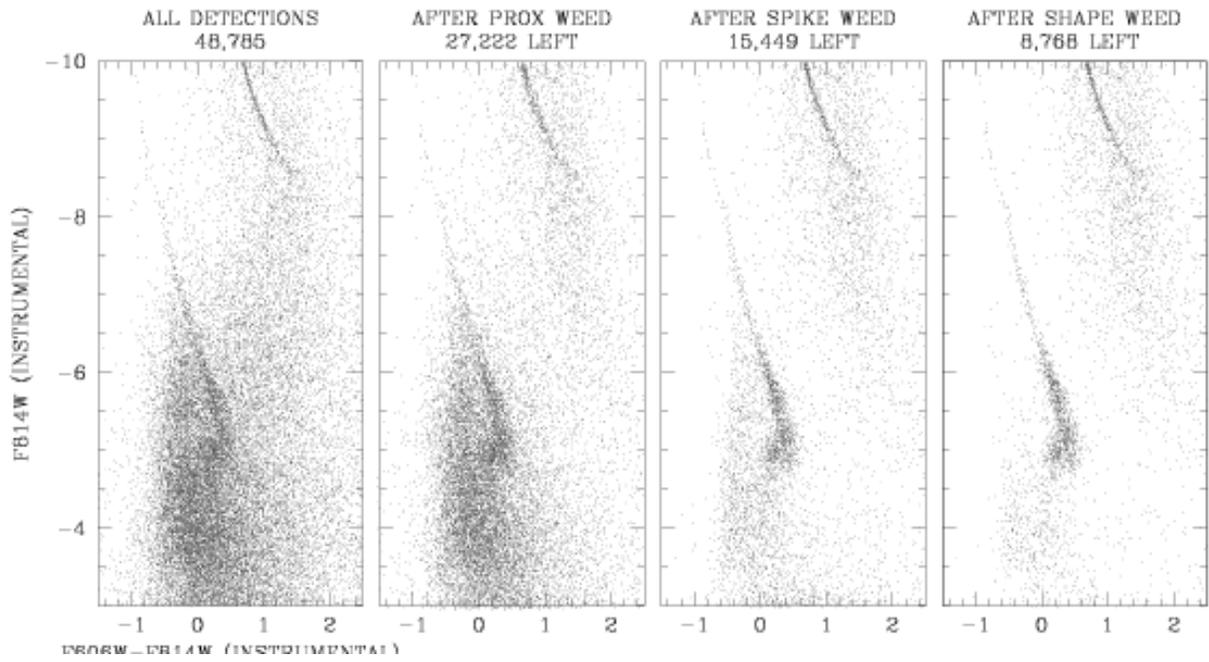}
\figcaption{This shows the progress of cleaning the CMD from the first
            weeding step to the last.
\label{FIG.CMDWEED}}
\end{figure}

Figure~\ref{FIG.CMDWEED} summarizes the weeding procedure.
Our final cut was to crop off the outer parts of the field, where
coverage was only partial because of our dithering; this excluded only a
small fraction of the sample.  We retained the quadrilateral region
defined by the points whose coordinates are (25,25), (4050,175),
(4150,4200), and (275,4000); this insured that the entire remaining area
had coverage by at least 242 of the 252 F814W images.  This left 8,399
stars.

\subsection{Possible CTE concerns}
\label{SSEC.CTE}
The charge-transfer efficiency (CTE) of the WFC has most recently 
been explored by Riess \& Mack (2004, RM04).  They evaluated the impact 
of charge-transfer inefficiency on a range of apertures, fluxes, and 
sky backgrounds.  It was not possible, though, for them to explore
the situation posed by our large data set, where the background 
is high and the target stars are only marginally above the background.
RM04 make the point that a high background can significantly mitigate 
CTE for faint stars.  This would make sense if faint stars are not much 
more likely to fill the charge traps than are the background pixels.
If this is the case, then CTE might not pose a significant problem.

This data set is not well devised to do a self-calibration of CTE 
contamination, since we have only one pointing at one orientation.  
The only way we have to evaluate whether CTE is affecting our faint 
stars is to see if the WD sequence appears brighter when there are 
fewer pixel transfers.  So, we divided the sample in two:  stars 
close to the readout registers ($y<1024$ or $y>3076$) and 
stars far from the registers ($y>1024$ and $y<3076$).   We compared
the two WD sequences and noticed that the truncation appeared at the
same F814W magnitude and color in both samples.  Thus, we conclude
that charge-transfer inefficiency does not significantly impact our
faint-star photometry.

\section{The final star list}
\label{SEC.FINALLIST}

The procedures that we have described were optimized for finding and
measuring the faintest stars, which are the main focus of this
investigation.  For brighter stars these procedures are not optimal,
however.  In this section, we will construct a single list that includes
both the faint stars and the bright stars.  We will also calibrate the
photometry.

\subsection{Merging the faint and bright lists}
\label{SSEC.MERGE}
Those stars that are bright enough to stand out clearly in each exposure 
are actually better measured by using PSF fitting in the individual
exposures, rather than by simultaneous fitting.  In \S~\ref{SUBSEC.TRANS} 
we measured average positions and fluxes for the bright stars.  These 
positions define the reference frame, and the photometry is zero-pointed 
to correspond to the instrumental system of the first exposure, the 
same system as used in \S~\ref{SEC.PHOT}.

Many stars in the field were bright enough to be saturated in the deep 
exposures, so we measured accurate fluxes for them in the short exposures.
Although the background in the deep exposures was high enough to make 
CTE concerns negligible (see \S~\ref{SSEC.CTE}), we would not expect this
to be the case for the shorter exposures.  Our measurements came from 
PSF fitting rather than aperture photometry, so none of the apertures 
evaluated by RM04 was appropriate to correct the CTE in our short exposures.
We thus had to develop our own specialized procedure for CTE corrections.  
What we did was to adopt the form of their Eq.\ (1), but to find an 
empirical value for the constants.  Since the sky in our images is always 
near the same value, we did not need to solve for a dependence on sky 
value, and thus had a dependence on only two unknowns, $\alpha$ and 
$\beta$: 
$$
 \Delta m_{\rm CTE} = \alpha (F_{1000})^{\beta} (y/2048),
$$
where $\alpha$ is a scaling factor and $\beta$ desrcibes the dependence 
on magnitude.  Here $F_{1000}$ is the flux of the star in the short 
exposures, in units of 1000 DN, and $y$ is the number of vertical 
pixel transfers.  The quantity $\alpha$ corresponds to the CTE suffered 
by a source at the far edge of the chip with a flux of 1000 DN.  For 
the 40-s short exposures, which had a background of 6 DN, we found 
$\alpha = 0.085$ and $\beta = -0.45$.  The dependence on magnitude 
($\beta$) is the same as given in RM04, but the scaling ($\alpha$) 
is about twice what they would predict for an aperture of 3 pixels'
radius (after a correction is made for the linear increase of CTE 
over time).  Since our aperture has an effective radius of much less 
than 3 pixels, it does make sense that we would suffer more CTE losses.  
In a similar way, we fitted curves to the difference between the 
photometry of the 40-s exposures and the 5-s exposures, and found 
$\alpha=0.09$ and $\beta = -0.45$.  This allowed us to correct the 5-s 
exposures for CTE.  Finally, we corrected the 1-s-exposure photometry
to the 5s-exposure photometry, finding $\alpha=0.1$ and $\beta = -0.75$.

Once we had corrected the photometry in all the short exposures for 
CTE, we then generated a single photometry list for the bright stars,
taking the photometry for each star from the deepest image where it was 
found to be unsaturated.  Only one star was bright enough to be 
saturated even in the 1-s exposures.  In this way, we arrived at a 
catalog of bright stars with the most accurate possible F606W and 
F814W instrumental magnitudes.

We now had two lists of stars.  One list, which we will call the ``faint'' 
list, came from the peak-based finding method, which found only the 
stars that were unsaturated in the deep exposures.  This list measures 
the faint stars accurately, but its photometry is not optimized for 
the stars with high signal-to-noise ratio.  This list is also the 
product of a carefully designed finding algorithm, which allows
us to evaluate its completeness accurately (see \S\ref{SEC.COMPL}).  
The other list (the ``bright'' list) has more accurate photometry 
for the bright stars, and contains all the stars brighter than about 
F814W magnitude $-7.5$ (1000 $e^-$), including all the stars that 
are saturated in the deep exposures.  

In order to arrive at a single list that combined the best qualities
of both lists, we needed to merge the bright and faint lists, which
overlap in the range of instrumental magnitudes brighter than $-7.5$
and fainter than $-13.5$ (where saturation in the deep images sets in).
We did the merging as follows:  We entered all of the stars from the
faint-star list into the final list.  Then for stars brighter than 
magnitude $-9$, where the magnitudes from the bright-star lists are 
more accurate, we substituted those in every case where they were 
available.  Finally, since the faint-star list does not include saturated 
stars, we added stars from the bright list that were brighter than 
$-13$ and were not already in the list.  This way, the membership in 
the list below $-13$ is entirely consistent with the faint-star finding 
criteria, so that a luminosity function constructed from the list can
be corrected for incompleteness by means of the artificial-star tests.  
Brighter than $-13$, the list is essentially 100\% complete, so no 
completeness corrections will be necessary.

\subsection{Photometric zero points}
\label{SSEC.ZP}
Thus far we have kept all of our photometry in the instrumental system,
since this system gives us an immediate handle on the signal-to-noise for
each source.  It is necessary, however, to determine the zero-point 
corrections that will convert these to actual astronomical magnitudes.  

Sirianni et al.\ (2005) find that for ACS, 1 count per second in a 
5\secspt5 aperture corresponds to magnitude 26.398\ in F606W and 25.501\ in 
F814W, in the VEGAMAG system (from their Table 10).  Recognizing, however, 
that 5\secspt5 is an impractically large aperture to use for actual 
measurements, they give corrections that convert to that aperture from 
the more practical aperture of 0\secspt5; their Table 5 gives for our 
two filters the additive quantities 0.088 and 0.087, respectively.  

\begin{figure}[!t]
\epsscale{1.00}
\plotone{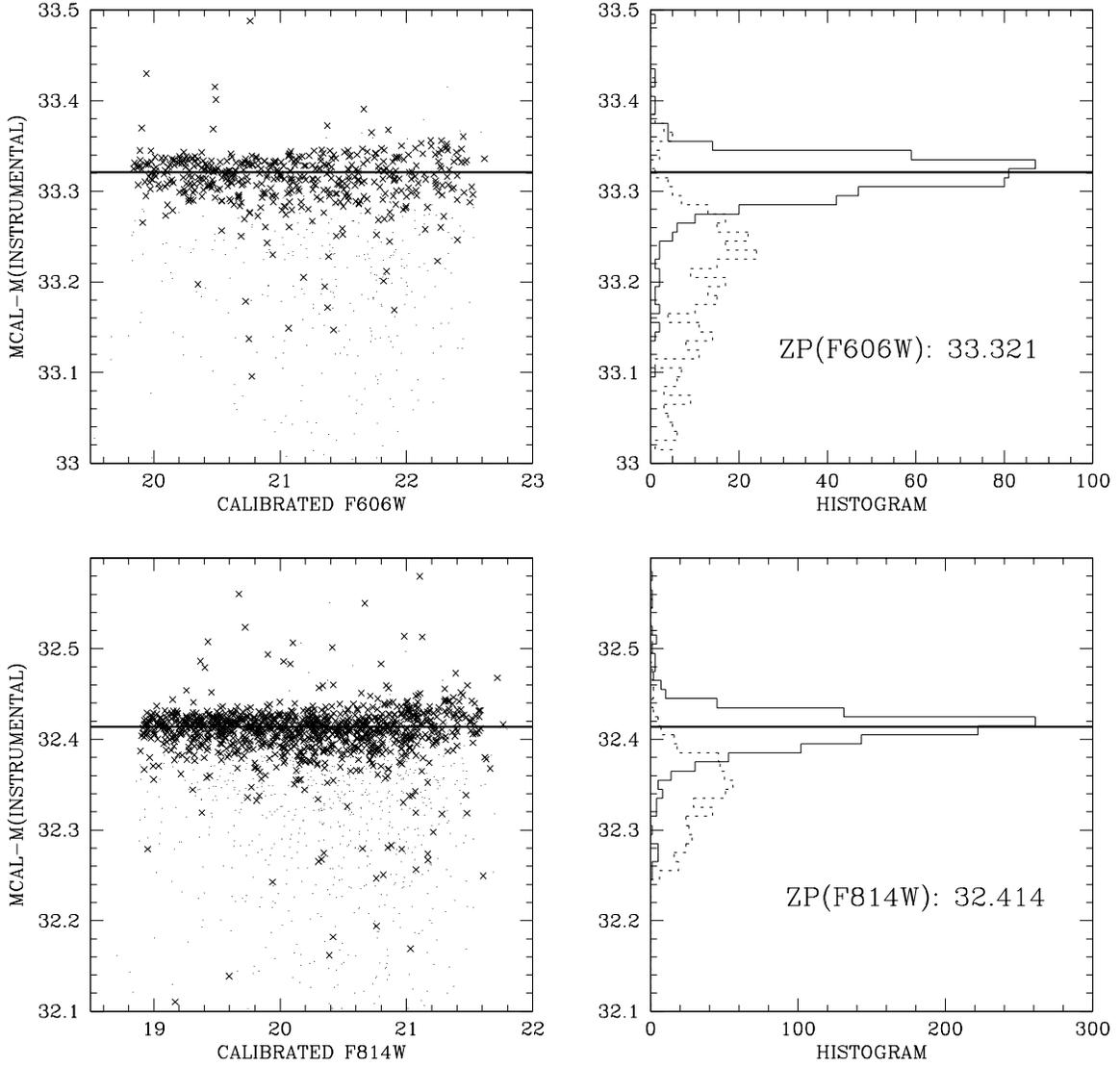}
\figcaption{({\it left}) The difference between our instrumental photometry and
                    the calibrated photometry for each star.  The points
                    represent stars that indicated additional contributions
                    to their aperture.  The adopted offset is shown as the
                    solid line. 
            ({\it right}) Histogram of photometric offset.  The dotted lines 
                    represent the compromised stars.
            Top plots are for F606W and the bottom plots for F814W.
\label{FIG.ZP}}
\end{figure}

Sirianni's calibrations were performed for the {\tt drz} images, so 
we used images {\tt j97101bdq\_drz} and {\tt j97101bbq\_drz} to construct 
calibrated magnitudes for isolated stars for F606W and F814W.
In Figure~\ref{FIG.ZP}, we compare the photometry for these calibrated 
stars against our instrumental photometry, to determine our zero point.
Even though this is a relatively sparse field, the 0\secspt5 aperture 
(10 pixels) often includes more than the target star (i.e., other faint 
stars or cosmic rays), so we identified stars that had inconsistent
fluxes in the $r=5$ and $r=10$ apertures and excluded them from the fit.
After this exclusion, we constructed a sigma-clipped average zero point 
for each filter.  We found our zero points to be 33.321 and 32.414 for 
F606W and F814W, respectively.  The histograms indicate that these should 
be good to about 0.01 magnitude.


\section{Completeness}
\label{SEC.COMPL}

Our goal in this study is to measure as many of the stars in the 
field as we can, as accurately as possible, and also to determine what
fraction of them we were able to measure.  Artificial stars helped 
with the finding and measuring approach, but for assessing the 
completeness they are crucial.  Section \ref{SUBSSEC.AS} describes 
how we set up our large grid of artificial-star tests, which covered 
the stars fainter than an F814W instrumental magnitude of $-11$.  
To extend the completeness values to brighter magnitudes, we generated
another set of images for the brighter stars, and joined the results of
finding and measuring these with the previous results for fainter stars,
so as to have completeness figures over our whole range of magnitudes.

The completeness of a catalog is a function both of the data set and of
the finding algorithm.  In \S~\ref{SUBSSEC.AS} we carefully inserted the
artificial stars into each individual exposure at the appropriate
position and with the appropriate flux, along with the proper noise.
The only difference between the real and artificial stars is that we
know the PSF and transformations exactly for the artificial stars,
whereas for the real stars we have good, yet still imperfect, models.
These small, unavoidable differences should have a negligible impact on
our completeness assessment, however.

\begin{figure}[!t]
\epsscale{1.00}
\plotone{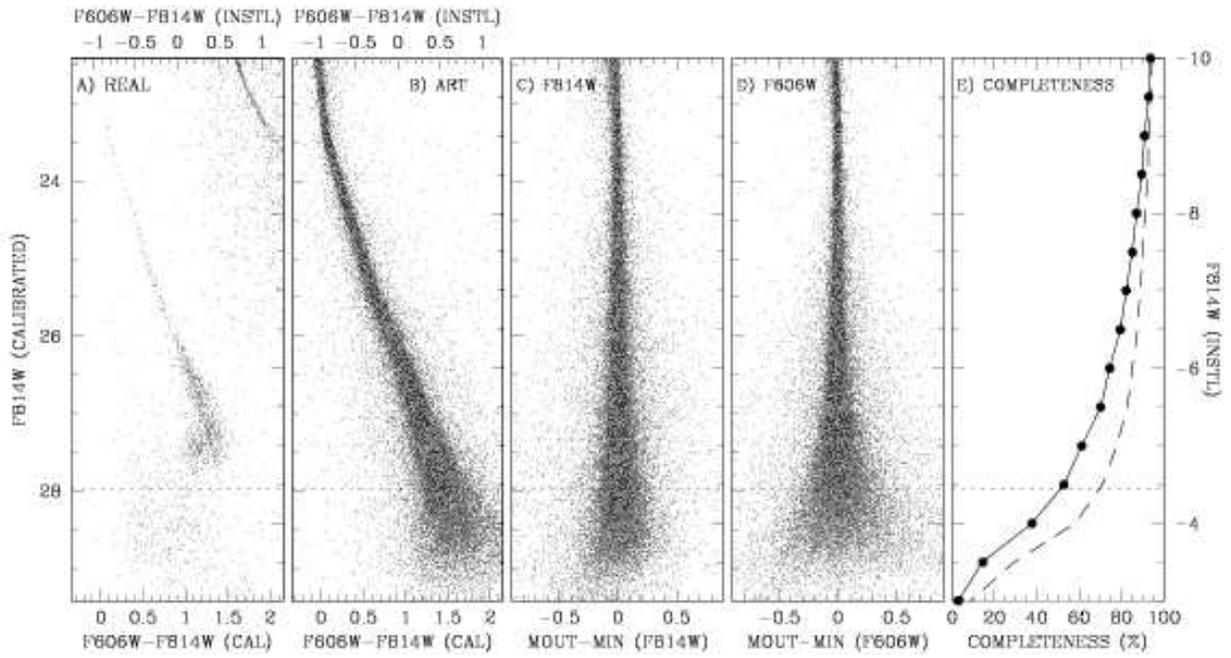}
\figcaption{The results of the artificial-star tests, all plotted
            against F814W.  
            (a) CMD of the real stars, for comparison; 
            (b) CMD of the artificial stars;
            (c) recovered photometry for F814W;
            (d) recovered photometry for F606W;
            (e) recovered completeness; the dotted line shows all
                stars that were recovered, while the solid line shows
                all recovered stars that also passed our weeding
                procedure.
\label{FIG.ART}}
\end{figure}

We subjected the AS images to the same finding and measuring procedure as 
the real stars.  This naturally includes all of the weeding-out operations
that we performed on the actual images.  Figure~\ref{FIG.ART} shows the CMD 
of the artificial stars that were inserted and that of the ASs that were 
recovered, along with plots showing the recovered magnitudes.  Even though 
the faintest stars were barely above the noise in individual exposures, 
our algorithm is able to measure fluxes for them that are systematically 
accurate.  The difference between the dashed line and the solid line in 
the right panel shows that there were clearly some stars that were found 
in the initial step, but did not pass the weeding scrutiny.  One might 
be tempted to relax the weeding criteria to include these stars, but any 
change in the weeding criteria would allow unreal artifacts to enter the 
real-star list.

To determine whether a particular artificial star was recovered, we 
cross-identified the input list with the output list and recorded the 
closest star in the output list to the input-list position.  If this star 
was within 0.75 pixel in $x$ and in $y$, and within 0.75 magnitude in F814W, 
we considered it found.  Hansen et al.\ (2007) provide detailed tables of 
the artificial-star-recovery statistics.

It is worth noting that even though the formal completeness is 50\% at
F814W $\sim$ 28, we are still able to recover stars almost down to F814W
= 29, where the completeness is below 20\%.  The reason for this is a
paradoxical property of the usual definition of completeness, (AS
recovered)/(AS inserted).  This definition is quite valid; if we use it,
we correctly derive the completeness figures by which the observed
numbers of stars need to be divided.  The paradox (if that is the right
term for it) is that completeness figures are often used as a criterion
of reliability too.  Here, however, that would not be appropriate.  The
reason is that in many parts of our field we fail to recover ASs not
because their faintness puts them near the limit of detection, but
rather because they fell in regions --- such as the surroundings of
bright stars --- where the background is just too noisy to give a faint
star a fair chance.  Thus we actually have reliable measurements of many
stars that are much fainter than the 50\%-completeness level.
Fortunately this distinction is not an issue for us in the present
paper, because all the sequences of interest terminate above the
50\%-completeness level.  In a different study in which some of us 
are involved, however, it plays a crucial role, and it is discussed in
greater detail there (Bedin et al.\ 2008).  (We note that in Bedin 
et al 2008, for NGC 6791, the WDCS does not terminate above the global 
50\% completeness line, and we take care to show that by restricting 
our searches for the faint stars only where the image is locally flat, 
we can maintain a completeness of greater than 50\% to a fainter 
magnitude.  This technique is not necessary for the present data set, 
but may be of benefit if we get a second ACS epoch for proper motions.)

\section{PROPER MOTIONS}
\label{SEC.PMs}

\begin{figure}[!t]
\epsscale{1.00}
\plotone{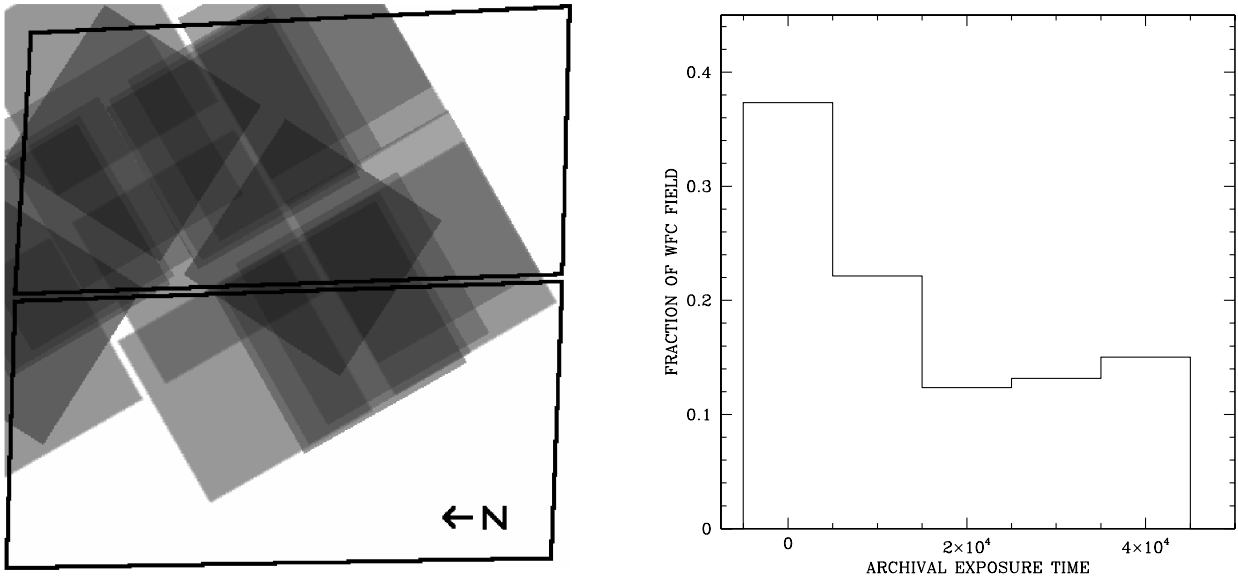}
\figcaption{({\it left})  Depth (in total exposure time) of the archival 
                          images as a function of position in our field.  
            ({\it right}) A histogram showing the fraction of our field with a 
                          given archival exposure time.
\label{FIG.ARCHIVE_DEPTH}}
\end{figure}

Our original plan for this project included a set of second-epoch
observations in the spring of 2007, which would give us a 2-year
baseline for proper motions, allowing us to separate the cluster stars
from field stars.  We had also chosen our present field to overlap 
existing archival images, so as to get some preliminary proper motions 
before the second ACS epoch.  With the failure of ACS in January 2007, 
however, we were restricted to using these archival images, taken from 
1994 to 1997, as our sole source of images from which to derive proper 
motions.  They do not go nearly as deep as we would wish, nor do they 
give uniform and complete coverage of our ACS field.  The archival data 
come from programs GO-5138, GO-5092, and GO-6797, with the exposure time 
divided about evenly between 1994 and 1997.  In total, there are 52 deep 
exposures, 35 in F814W (33,000 s) and 17 in F606W or F555W (16,500 s).  
Figure~\ref{FIG.ARCHIVE_DEPTH} shows the coverage depth.  We have a 
different limiting archival magnitude in different parts of the field, 
making each star a special circumstance and making completeness an 
impossible question to answer for the archival data.

Worse, even where it exists, our archival coverage is pitifully shallow.
Our ACS exposures total 180,000 sec in F814W and 93,000 sec in F606W.
When the $\times$4 greater sensitivity of ACS is a also taken into
consideration, the total exposure with the ACS WFC is much more than an
order of magnitude longer than the deepest archival coverage.  We do
have one advantage, however, that partly offsets the lack of depth in
the archival material:\ our ACS star list tells us where to look for
stars in the earlier WFPC2 images.  Unfortunately this advantage comes
with a limitation.  The cluster stars all have the same motion (except
for the very small differences caused by internal motions), so that once
we have established the transformations, we know exactly where to look
for each cluster star.  The field stars, by contrast, have a considerable 
scatter about their mean proper motion, so finding them is less certain.

Our first step in dealing with the archival images was to establish
transformations between each chip in each of the archival exposures and 
our ACS WFC reference frame.  For this purpose we identified in the WFC
lists 3,510 bright stars that fall along the cluster sequence in the CMD
and also fell in the regions covered by the archival images.  We derived
the transformations much as we had done for our WFC images (\S\
\ref{SUBSEC.TRANS}), except that for the archival images we naturally
used the distortion corrections appropriate for WFPC2.

In deriving the transformations from cluster stars alone, we had
guaranteed that all cluster members would be found in the positions that
the transformations predicted for them, since they have the same motion
as the stars from which the transformations were derived (again
neglecting the internal motions as being too small to matter).
Measuring them is, except for faintness, no worse a problem than was
measuring these stars in the ACS images.

\begin{figure}[!t]
\epsscale{0.90}
\plotone{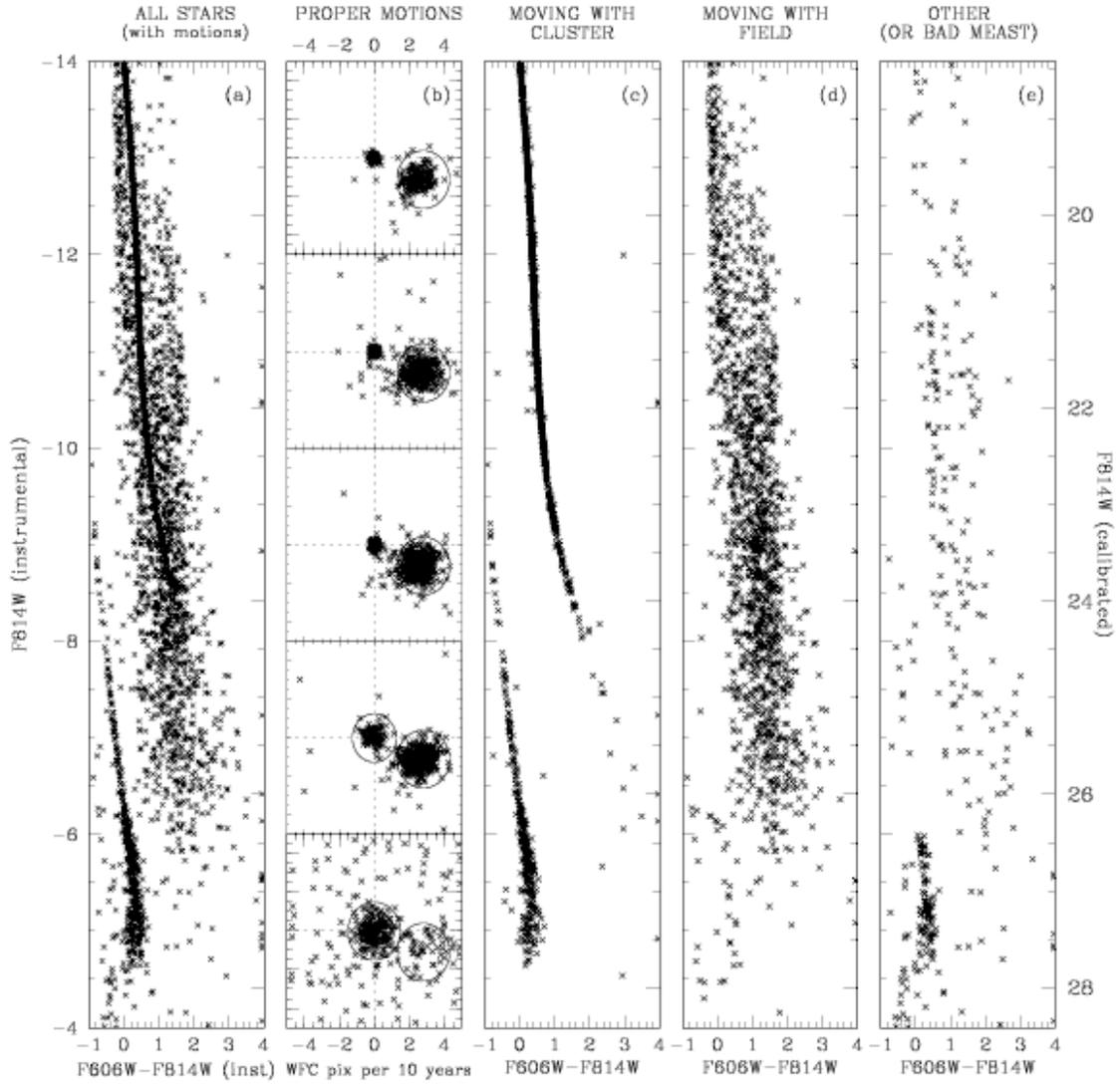}
\figcaption{ (a) The color magnitude diagram for all stars.  Stars redder 
                 than an instrumental color of $4$ are plotted on the right 
                 edge of the diagram.  
             (b) The proper-motion diagram for stars in bins 2 magnitudes 
                 tall.  Stars moving with the cluster should be at the 
                 origin.  Circles show the regions used to separate members
                 and field stars.  
             (c) The CMD for stars that are moving with the cluster.
             (d) The CMD for stars moving with the field.  
             (e) The CMD for stars not clearly moving with either
                 (probably due to large non-cluster motions, or to large
                 motion errors ).
\label{FIG.CMDnPM}}
\end{figure}

Field stars are a different matter, however.  Not only do they have a
different mean motion; they also have a large scatter about the mean.
The difference in motions is shown in Figure \ref{FIG.CMDnPM}.
King et al.\ 1998 show an analogous figure for the same field, but here 
we have a much longer baseline for the motions:  10 years versus 3 years.
Since we had no way of knowing whether any given faint star
was a member of the cluster or the field, we did as follows: We started
by making a list of every single peak (local maximum) in each of the
WFPC2 exposures.  For each star we collected all the peaks that were
within 5 pixels of the place where the star would fall if it were a
cluster member.  If the star does not have the cluster motion, its 1994
and 1997 peaks should have locations that differ systematically,
however.  To allow for this, we normalized the displacement of every
peak to a 10-year baseline:\ each peak in a 1994 image was moved toward
the cluster-star position by 1/11th of its displacement, and each peak
in a 1997 image was moved away from the cluster-star position by 2/8ths
of its displacement.  If there was a significant clumping of peaks, we
took its centroid as the 10-year displacement of the star.  (The 
8-year/11-year adjustment was made for all stars, including cluster 
members, for which it of course made almost no difference.)

Our measured proper motions are shown in panel (b) of
Fig.~\ref{FIG.CMDnPM}, in which their distributions are plotted
separately for successive intervals of 2 magnitudes.  Circles around the
cluster and field clumps show our selection criteria for cluster and
field.  The other four panels of the figure are CMDs:\ in (a) all the
stars, in (c) the cluster stars, in (d) the field stars, and in (e) the
stars that did not fall within the circle for either group.  It is clear
that in the brighter four magnitude bins our cluster/field separation is
fairly good.  In the faintest bin, by contrast, the separation is quite
poor.

This inability to make a proper-motion separation for the faintest stars
properly has several effects.  For the main-sequence stars it hardly
matters, since their number has dropped virtually to zero by the time we
reach that magnitude.  For the white dwarfs, the clear presence of the
faint end of their sequence in the ``other'' CMD, and their faint
presence in the ``field'' CMD, show that proper motions cannot be used 
reliably to establish membership fainter than F814W magnitude 26.5.

The CMD of the field stars is more complicated.  In the brightest part
of panel (d) is a sharp blue edge, which must come from stars near the
main-sequence turn-off, at a range of distances in the bulge/halo
population of the background.  The redder stars in that CMD are probably
a mixture of the background population and the red-dwarf foreground.  The
latter is probably exemplified by most of the magnitude range in panel
(e), since the foreground stars can be expected to have larger proper
motions.  Furthermore, the sparseness of stars in panel (e) suggests
that this population makes only a small contribution to panel (d),
whereas it continues in panel (e) because inaccuracy of proper motions
does not remove stars from this category.

Interestingly, the star numbers in the field-star panel, (d), peter out
between F814W magnitudes 26 and 27, and there are almost none of them
fainter, but no corresponding star density appears in panel (e), as it
would if their dearth in panel (d) were due to the increased measurement
error for fainter stars.  Since white dwarfs are plentiful down nearly
to magnitude 28 in the sum of panels (c), (d) and (e), or what is the
same thing, in panel (a), we can only conclude that the background
population really is lacking at magnitudes fainter than 27.

\section{Summary of data products, published papers, and papers to come}

This paper describes how we started with a dithered set of 378 deep 
exposures taken in the outskirts of the globular cluster NGC 6397 over 
126 HST orbits, and produced the most comprehensive possible list 
of the point sources in the field.   The faintest stars could not be
independently identified in every image, so we collated information 
from all of the images and identified a possible star wherever a 
statistically significant number of images showed a minimal detection 
(a simple peak, or local maximum).  This list contained all the identifiable 
stars, but also contained PSF features, diffraction spikes, and galaxies.
(In fact, this list of detections also contained some globular clusters 
surrounding a distant background galaxy, see Kalirai et al.\ 2008.)
We measured each of these objects under the presumption that it is a 
star, determining a position, a flux in F606W and F814W, and some other 
diagnostics.  We then subjected this list to a set of criteria designed 
to leave only the point sources.  These criteria were able successfully 
to reveal a white-dwarf sequence that increases in number toward faint 
magnitudes, turns to the blue, then appears to truncate at F814W = 27.5  
at a color of F606W $-$ F814W $\sim$ 1.2.  Artificial-star tests showed 
that our reduction procedures could have easily found stars a magnitude 
fainter than the observed truncation.

We also made use of data in the archive to measure proper motions
for as many stars in the field as possible.  We effected a near-perfect
separation for stars down to the end of the main sequence, but the
proper motions were useful only in a qualitative sense for the white
dwarfs.  

We have thus determined a position and a F606W and F814W flux for all
stars, and proper motions for most.  In addition, we have also performed
artificial-star tests to show the expected photometric errors and
completeness.  These data products are the subject of several papers:\
Richer et al.\ (2006) presented the initial discovery of the 
termination of the WD sequence and of the hook to the blue.
Kalirai et al.\ (2007) have determined the absolute motion of the
cluster by studying the relative motions of cluster stars and background
galaxies.  Hansen et al.\ (2007) analyze the white-dwarf population 
in detail.  Davis et al.\ (2007) examine the spatial distribution of 
the WDs in this cluster.  In companion papers in this volume, Richer 
et al.\ (2008) explore the bottom of NGC397's main sequence and Hurley 
et al.\ (2008) discuss the dynamical state of the cluser.  Finally, 
Kalirai et al.\ (2008) study the globular cluster system of a background 
galaxy seen in this field.

The only disappointing aspect of this reduction is the quality of the 
faint-star proper motions, which is limited by the shallow and inhomogeneous 
archival images.  If ACS is restored in Servicing Mission 4 in 2008,
then with second-epoch observations we will be able to determine 
membership for all stars, and even measure internal motions for the 
cluster stars.

\acknowledgements

All the US Co-I's (J.A.,B.M.S.H, I.R.K., J.K., R.M.R., and
M.S.) acknowledge the support of HST grant GO-10424.  The research 
of H.B.R. is supported in part by the Natural Sciences and Engineering 
Research Council (NSERC) of Canada.

\clearpage 

\references

Anderson, J,. \& King, I. R. 2000, PASP, 112, 1360

Anderson, J., \& King, I. R. 2003, PASP, 115, 113

Anderson, J., \& King, I. R. 2006. ACS ISR 2006-01 (Baltimore:\ STScI)

Bedin, L. R., King, I. R., Anderson, J., Piotto, G., Salaris, M., Cassisi, S., \&
   Serenelli, A.  2008.  ApJ, in press.

Davis, D. S., Richer, H. B., King, I. R., Anderson, J., Coffey, J., Fahlman, G. G.,
   Hurley, J., \& Kalirai, J.  2008, MNRAS 383 L20

Fruchter, A. S., \& Hook, R. N. 2002, PASP, 114, 144

Hansen, B. M. S., et al.  2007 ApJ 671 380

Hurley, J., et al. 2008, AJ, submited

Kalirai, J., et al. 2007, ApJ, 657, L93

Kalirai, J., et al. 2008, submitted to ApJ

Krist, J. 2003, ACS ISR 2003-06 (Baltimore: STScI)

Lauer, T. R. 1999, PASP, 111, 227L

McLaughlin, D. E., Anderson, J., Meylan, G., Gebhardt, K., Pryor, C., Phinney, S.
     2006, ApJS, 166, 249

Pavlovsky, C., et al. 2004, "ACS Data Handbook," Version 3.0, (Baltimore: STScI)

Richer, H. B., et al.  1997, ApJ, 484, 741 

Richer, H. B., et al.  2004, AJ, 127, 2904

Richer, H. B., et al.  2006, Science, 313, 936

Richer, H. B., et al.  2008, AJ, in this volume

Riess, A., \& Mack, J. 2004, ACS ISR 2004-06 (Baltimore: STScI)

Sirianni, M., et al. 2005, PASP, 117, 1049

Stetson, P. B. 1987, PASP, 99, 191

\clearpage 

\clearpage 

\clearpage 

\clearpage 

\clearpage 

\clearpage 

\clearpage 

\clearpage 

\clearpage 

\clearpage 

\clearpage 

\clearpage 

\clearpage 

\clearpage 

\end{document}